# Accuracy of Flight Altitude Measured with Low-Cost GNSS, Radar and Barometer Sensors: Implications for Airborne Radiometric Surveys

Matteo Albéri [1,2,*], Marica Baldoncini [1,2], Carlo Bottardi [1,2], Enrico Chiarelli [3], Giovanni Fiorentini [1,2], Kassandra Giulia Cristina Raptis [3], Eugenio Realini [4], Mirko Reguzzoni [5], Lorenzo Rossi [5], Daniele Sampietro [4], Virginia Strati [3] and Fabio Mantovani [1,2]

[1] Department of Physics and Earth Sciences, University of Ferrara, Via Saragat, 1, 44122 Ferrara, Italy; baldoncini@fe.infn.it (M.B.); bottardi@fe.infn.it (C.B.); fiorenti@fe.infn.it (G.F.); mantovani@fe.infn.it (F.M.)
[2] Ferrara Section of the National Institute of Nuclear Physics, Via Saragat, 1, 44122 Ferrara, Italy
[3] Legnaro National Laboratory, National Institute of Nuclear Physics, Via dell'Università 2, 35020 Legnaro (Padova), Italy; enrico.chiarelli@student.unife.it (E.C.); kassandragiul.raptis@student.unife.it (K.G.C.R.); strati@fe.infn.it (V.S.)
[4] Geomatics Research & Development (GReD) srl, Via Cavour 2, 22074 Lomazzo (Como), Italy; eugenio.realini@g-red.eu (E.R.); daniele.sampietro@polimi.it (D.S.)
[5] Department of Civil and Environmental Engineering (DICA), Polytechnic of Milan, Piazza Leonardo da Vinci 32, 20133 Milano, Italy; mirko.reguzzoni@polimi.it (M.R.); lorenzo1.rossi@polimi.it (L.R.)
* Correspondence: alberi@fe.infn.it; Tel.: +39-329-0715328

**Abstract:** Flight height is a fundamental parameter for correcting the gamma signal produced by terrestrial radionuclides measured during airborne surveys. The frontiers of radiometric measurements with UAV require light and accurate altimeters flying at some 10 m from the ground. We equipped an aircraft with seven altimetric sensors (three low-cost GNSS receivers, one inertial measurement unit, one radar altimeter and two barometers) and analyzed ~3 h of data collected over the sea in the (35–2194) m altitude range. At low altitudes (H < 70 m) radar and barometric altimeters provide the best performances, while GNSS data are used only for barometer calibration as they are affected by a large noise due to the multipath from the sea. The ~1 m median standard deviation at 50 m altitude affects the estimation of the ground radioisotope abundances with an uncertainty less than 1.3%. The GNSS double-difference post-processing enhanced significantly the data quality for H > 80 m in terms of both altitude median standard deviation and agreement between the reconstructed and measured GPS antennas distances. Flying at 100 m the estimated uncertainty on the ground total activity due to the uncertainty on the flight height is of the order of 2%.

## 1. Introduction

Airborne Gamma-Ray Spectroscopy (AGRS) is a proximal remote sensing method that allows quantifying the abundances of natural ($^{40}$K, $^{238}$U, $^{232}$Th) and artificial (e.g., $^{137}$Cs) radionuclides present in the topsoil (~30 cm depth) over relatively large scales. Studying the spatial distribution of these radionuclides is strategic for monitoring environmental radioactivity [1], producing thematic maps of geochemical interest [2–4], identifying radioactive orphan sources [5] or investigating areas potentially contaminated by nuclear fallout [6]. Sodium iodide scintillation detectors (NaI(Tl)) are widely employed in AGRS measurements thanks to the high portability and high detection efficiency which allow performing surveys over extended areas in reasonable times and minimizing costs.

In the last decade there has been a strong effort in improving spectral analysis techniques which led not only to high-accuracy identification of radionuclides present in the environment [7], but also to the possibility of performing real time surveys, especially in the framework of homeland security applications [8]. The spread of Unmanned Aerial Vehicles (UAVs) is boosting research and development in the field of AGRS applied to innovative sectors, such as precision farming or emergency response in case of nuclear accidents, both in terms of the technologies employed (e.g., prototypes equipped with CdZnTe detectors [9]) and of spectral analysis methods [10]. UAVs equipped with a lightweight gamma spectrometer typically fly at altitudes and speeds lower than a helicopter, with the objective of enhancing the terrestrial gamma signal intensity [11].

In these new technological scenarios a precise evaluation of flight altitude is mandatory for avoiding systematic effects in the gamma signal corrections. In the last decade sophisticated analytical techniques based on inverse problem methods [12] as well as Monte Carlo simulations [13] have been proposed for studying Digital Elevation Model (DEM) effect corrections together with corresponding uncertainties in airborne gamma-ray spectrometry. This work addresses these topics improving the analysis of data collected from seven altimeters and reducing a source of uncertainty like DEM. In particular altitude measurements have been performed with four low-cost GNSS receivers, one radar altimeter and two low-cost barometric sensors in a series of flights over the sea exploring a wide range of altitudes (from 35 to 2194 m; Table 1 and Figure 1). The goal of this paper was to estimate the accuracies of flight altitude, investigating statistical and systematic effects due to calibration methods, post-processing analysis and sensor performances.

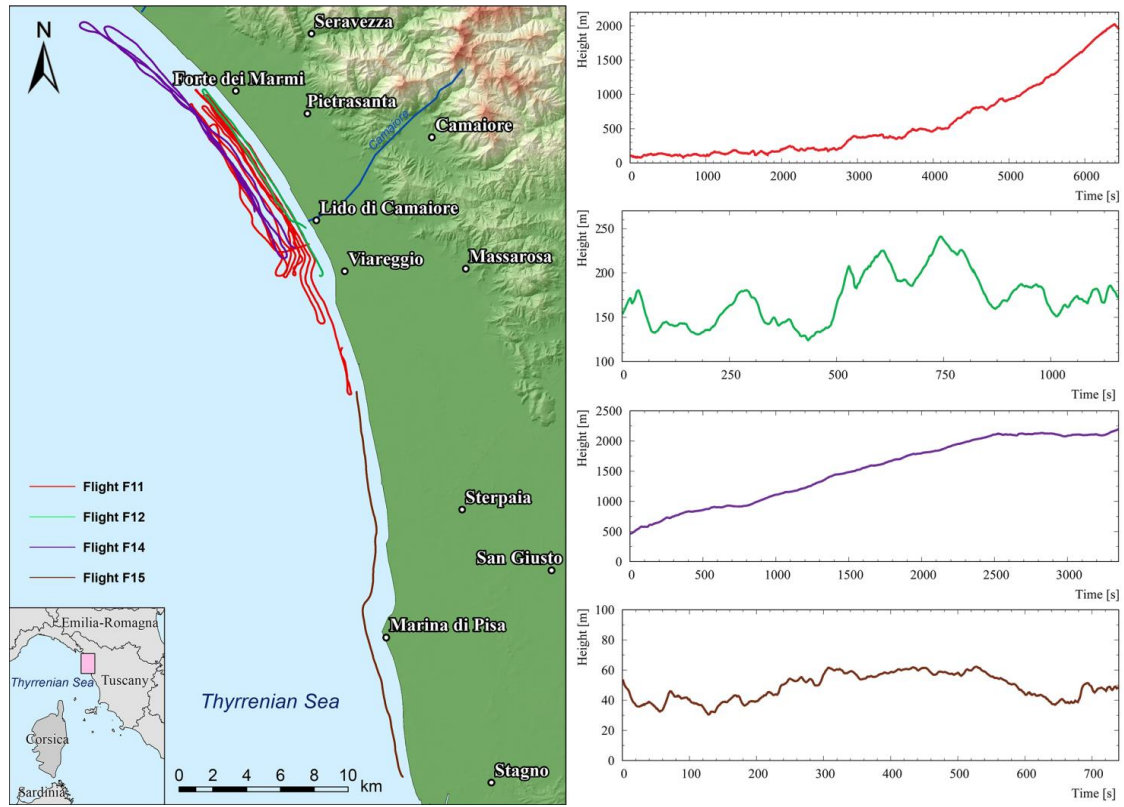

**Figure 1.** The left panel shows a map of the paths flown during the four surveys over the sea between Forte dei Marmi (LU) and Marina di Pisa (PI) in Italy. The four panels on the right show the mean altitude profiles measured by GPSABC of each flight.

**Table 1.** Main parameters of the four flights. $H_{min}$ and $H_{max}$ (minimum and maximum height) refer to the flight height above sea level calculated by averaging the measurements of the different sensors. Average horizontal and vertical speeds are calculated using the data from GPSABC.

| Flight ID | Date | Time (CEST) | $H_{min}$ (m) | $H_{max}$ (m) | Acquisition Time (s) | Average Horizontal Speed (m/s) | Average Vertical Speed (m/s) |
|---|---|---|---|---|---|---|---|
| F11 | 30/03/16 | 17:42:11–19:29:38 | 79 | 2018 | 6447 | 18.9 | 0.8 |
| F12 | 31/03/16 | 18:13:55–18:33:12 | 129 | 237 | 1158 | 15.5 | 0.5 |
| F14 | 05/04/16 | 16:37:15–17:33:04 | 464 | 2194 | 3350 | 21.1 | 0.8 |
| F15 | 05/04/16 | 19:15:19–19:27:39 | 35 | 66 | 740 | 34.4 | 0.6 |

## 2. Instruments and Methods

The aircraft used for the surveys is the Radgyro (Figure 2–4), an experimental autogyro devoted to airborne multiparametric measurements, specifically designed for radiometric surveying [14,15]. The Radgyro is 5.20 m long and 2.8 m high, and has an 83-liter fuel tank placed above the instrumentation to avoid the attenuation of gamma signals coming from the ground due to the interaction with the fuel material. The fuselage has been modified to house the experimental setup for an overall instrumental payload capacity of 120 kg which corresponds to a flight autonomy of approximately 3 h. Moreover, the Radgyro has two lateral aerodynamic compartments hosting infrared, thermal and visible cameras and an Inertial Measurement Unit (IMU). The Radgyro needs air through its rotor to generate lift so it cannot hover or take off vertically.

The detector used for gamma spectroscopy measurements is accommodated inside the main compartment of the Radgyro. It consists of four $10 \times 10 \times 40$ cm sodium iodide scintillation detectors (NaI(Tl)) for a total detection volume of about 16 liters [16]. The high payload and autonomy, the modularity of the acquisition system, along with the possibility of synchronizing measurements coming from different sensors with respect to the reference computer time, make the Radgyro a promising prototype aircraft for exploring new applications in the field of proximal remote sensing.

The Radgyro is equipped with seven altimetric sensors, belonging to three different instrumental classes: four GNSS antennas (GPSABC, GPSIMU), two pressure and temperature sensors (PT and PTIMU) and one radar altimeter (ALT) (Figure 3). In this study the height of Radgyro is referred to the GNSS antenna locations, which are located at the same vertical position with respect to the ground (1.08 ± 0.01 m). Taking into account that the radar altimeter accuracy is of the order of 3% of the measured altitude (see Section 2.2), the difference in distance from the ground between ALT and GNSS antennas (0.71 m) is negligible. The pressure sensors are calibrated using the GNSS (see Section 2.4) and therefore the barometric altitude is referred to GNSS antenna position.

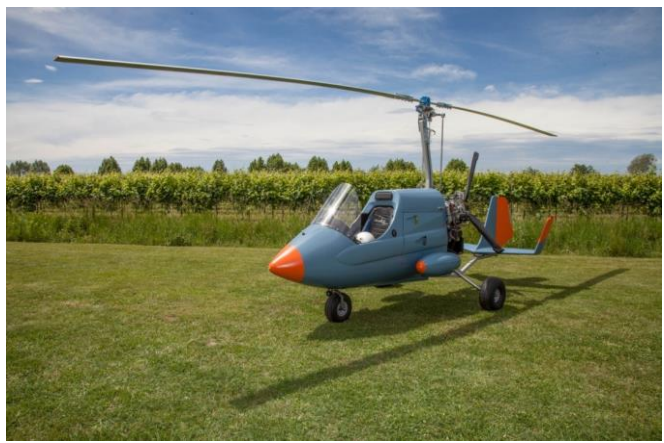

**Figure 2.** Radgyro, the autogyro used for all the surveys described in Table 1.

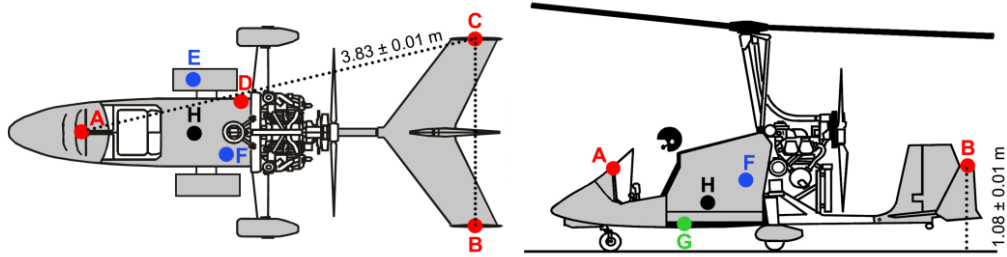

**Figure 3.** Scheme of the placement of the different devices on the Radgyro: (A) GNSS antenna (GPSA), (B) GNSS antenna (GPSB), (C) GNSS antenna (GPSC), (D) GNSS antenna connected to IMU (GPSIMU), (E) pressure and temperature sensors of IMU (PTIMU), (F) pressure and temperature sensors (PT), (G) radar altimeter (ALT), (H) gamma spectrometer NaI(Tl). GPSA, GPSB and GPSC are placed at the following relative distances: dGPSAB = dGPSAC = (3.83 ± 0.01) m and dGPSBC = (1.96 ± 0.01) m.

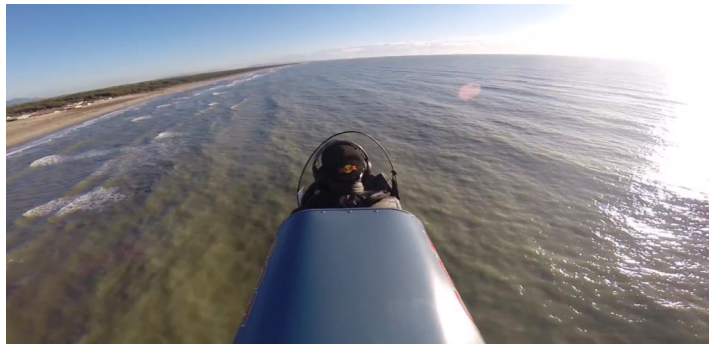

**Figure 4.** A typical situation of flight over the sea with the Radgyro.

*2.1. The Inertial Measurement Unit*

The right lateral compartment of the Radgyro houses the MTi-G-700 GPS/INS Inertial Measurement Unit (IMU, Figure 3), which is equipped with a GNSS receiver acquiring the GPS signal with a maximum frequency of 4 Hz, a barometer providing the atmospheric pressure readout with a maximum frequency of 50 Hz (PTIMU) (see Section 2.4) and inertial sensors retrieving the roll, pitch and yaw angles with a maximum frequency of 400 Hz. The IMU provides height values by combining the data from the GNSS, the barometer and the accelerometers with a maximum frequency of 400 Hz (GPSIMU). Dynamic and barometric measurements allow for height estimation even with weak GNSS signal and the nominal accuracy on the vertical position is 2 m (1σ) [17].

*2.2. The Radar Altimeter*

The Smartmicro® Micro Radar Altimeter (ALT), placed under the Radgyro fuselage (Figure 3), measures the flight altitude at ~60 Hz by using a radar sensor operating at a frequency of 24 GHz. The estimate of the minimum distance is declared reliable within a cone having 20° opening angle and the declared accuracy on altimetric measurements is 3%, with a minimum value of 0.5 m. Although the flight altitude range declared by the seller is (0.5–500) m, our data analysis on the ALT dataset revealed a significative presence of outliers at heights above 340 m (Figure 5). Neglecting effects related to wave motions and tidal variations, which are typically <0.4 m in the surveyed area [18], we performed our study considering two different datasets named α and β, corresponding respectively to H < 340 m and H > 340 m respectively. The α database is populated by data acquired in 4803 s by all 7 sensors, while the β database refers to the remaining 6892 s in which the ALT sensor is excluded (Table 2).

**Table 2.** Number of entries of the datasets used in the analysis of the orthometric height values from GNSS, altimetric and barometric measurements. The 340 m height cutoff has been identified on the base of altimeter outlier data (see Figure 5), while the 0.2 Hz frequency is related to the availability of the Madonna Dell'Acqua master station data for the GNSS post-processing.

| Datasets | Frequency | α<br>H < 340 m | β<br>H > 340 m |
|---|---|---|---|
| DATASET 1 | 1.0 Hz (stand-alone) | 4803 | 6892 |
| DATASET 2 | 0.2 Hz (double-difference) | 960 | 1378 |

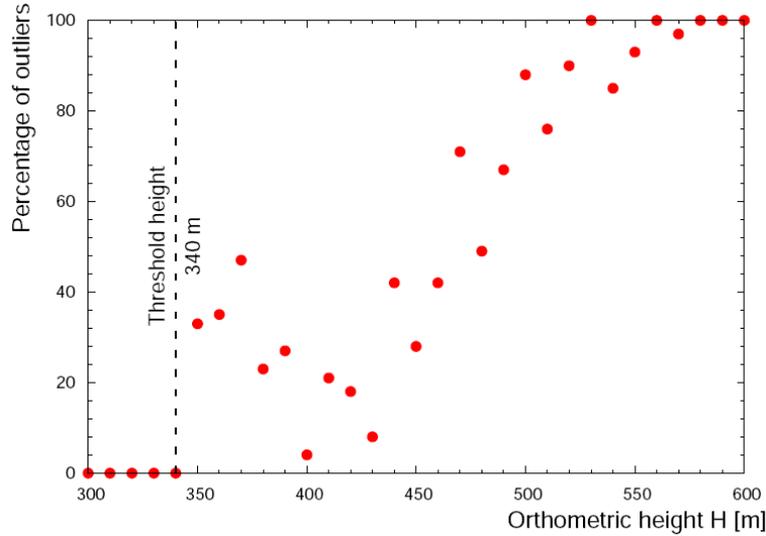

**Figure 5.** Percentage of outliers in the ALT dataset as a function of the orthometric height. The altitude of 340 m has been identified has a threshold above which the ALT dataset has been excluded from the global analysis

*2.3. The Three GNSS Receivers*

The Radgyro is equipped with three single frequency u-blox EVK-6T receivers, each of them coupled with a GPS ANN-MS active antenna having ~100 g weight and dimensions of 48 mm × 40 mm × 13 mm, one located on the aircraft cockpit (GPSA) and the others on the tail wings (GPSB and GPSC) (Figure 3). A low noise amplifier is implemented on each receiver which is intended to compensate the loss of signal due to cables and connectors. The u-blox EVK-6T receivers contain LEA-6T modules, that can provide raw GPS data as output, allowing for advanced post-processing. The cost of an EVK-6T receiver and ANN-MS antenna is of few hundred Euros.

Each GPS receiver is able to directly deliver a real-time solution, using NMEA GGA sentences, and raw data to be post-processed in standard RINEX format, both with a sampling frequency of 1 Hz. Moreover, the logging software records the GPS acquisition time coupled with the absolute computer time in order to correctly synchronize GPS with the other sensors present onboard. GPS raw observations were post-processed following two different analyses with the open source goGPS software [19]:

- code-only stand-alone solution (1 Hz), using a Kalman filter with constant-velocity dynamics;
- code and phase double differences solution (0.2 Hz) with respect to the permanent station Madonna Dell'Acqua (Pisa) (43.7475° N, 10.3660° E, 2 a.s.l), using a Kalman filter with constant-velocity dynamics.

These different methods produce two datasets that we define as DATASET 1 and DATASET 2 at 1 Hz and 0.2 Hz respectively (Table 2). Regarding GNSS-derived heights, the EGM2008 model [20] was used to convert ellipsoidal heights to orthometric ones, since it is the model currently implemented in the goGPS software.

The identification of GNSS data outliers has been performed by studying the distribution of the distances reconstructed between the three GPS antennas $d_{GPSAB}$, $d_{GPSAC}$, $d_{GPSBC}$ with respect to reference values (Figure 3). Following [21], an outlier is a data point that lies out of the ranges ($Q_1 - 1.5$ IQR) and ($Q_3 + 1.5$ IQR), where $Q_1$, $Q_3$ and IQR are first quartile, third quartile and interquartile range respectively. Outlier data have been typically recognized when flying close to the sea (Figure 6) and at an altitude range of (35–900) m (Figure 7 and Figure **8**). The analysis of outlier highlights that their percentage generally decreases with increasing altitude and that the median $d_{GPSBC}$, $d_{GPSAC}$ and $d_{GPSAB}$ approach the reference distances.

We note that in F15 the $d_{GPSAB}$ erraticity decreases drastically crossing the border between sea and land (Figure 6). The average reconstructed $d_{GPSAB}$ varies from (5.86 ± 7.18) m (over water) to (3.77 ± 0.28) m (over land), to be compared with the (3.83 ± 0.01) m reference distance. In F15, characterized by a (35–66) m flight altitude range, it is possible to observe a noise amplification due to the multipath effect over the sea. This phenomenon is well known in literature and has been studied in different environmental scenarios [22], investigating also applications like the monitoring of coastal sea levels and of the periodicity of ocean tides [23–25].

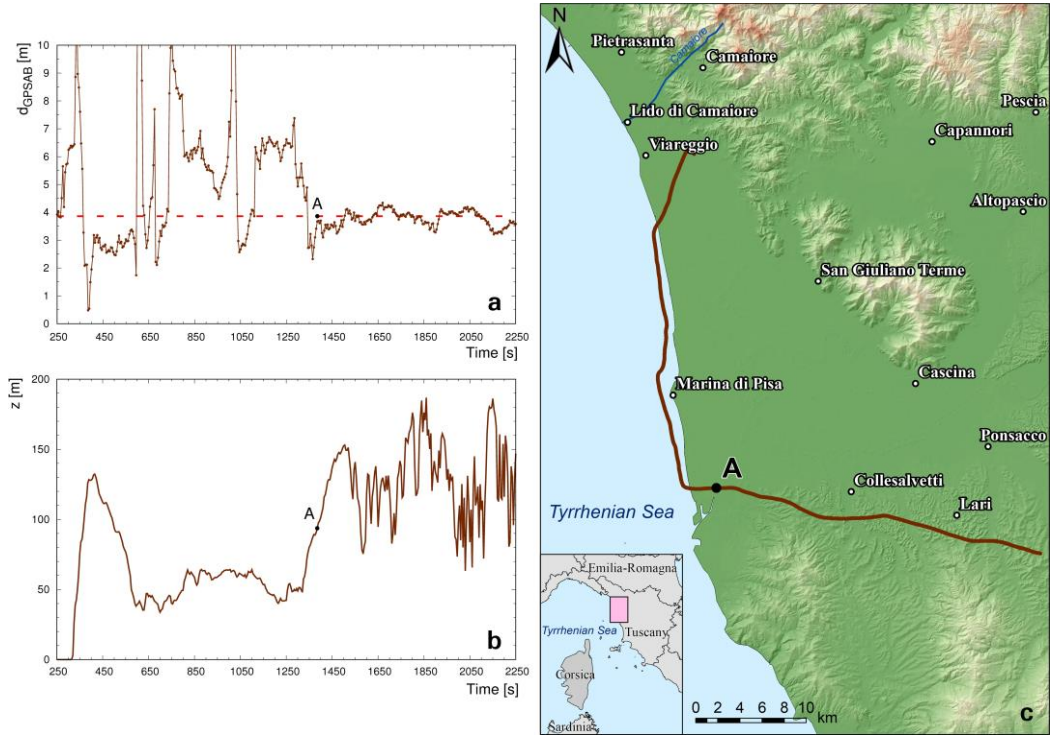

**Figure 6.** (**a**) Reconstructed distance between GPSA and GPSB as a function of time during a portion of F15. The dashed red line represents the (3.83 ± 0.01) m reference distance and the brown line represents the average reconstructed $d_{GPSAB}$ during the flight. The large fluctuations observed in the reconstructed distance when flying over the sea are strongly reduced when flying over land, in particular when flying more than 3 km far from the coast (point A). (**b**) Mean height above the ground level z(m) (digital elevation model is subtracted) measured by GPSABC.

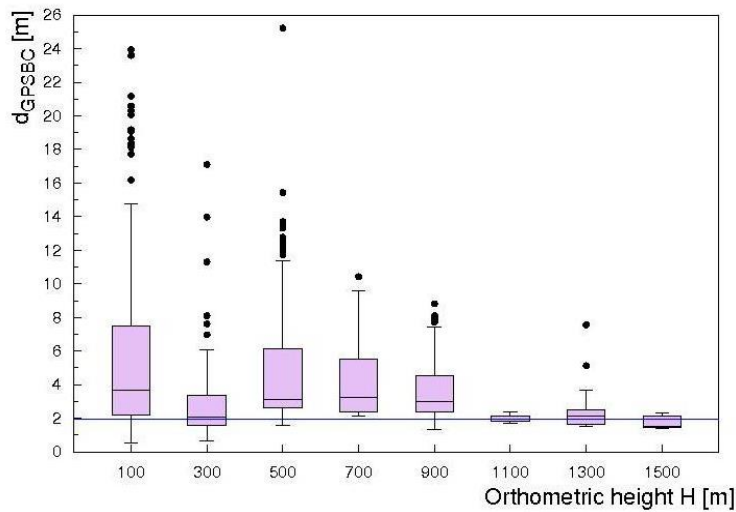

**Figure 7.** Boxplot of the distribution of $d_{GPSBC}$ as a function of the orthometric height H for entire 0.2 Hz dataset. The blue line represents the (1.96 ± 0.01) m reference distance between GPSB and GPSC. Black points represent outlier data.

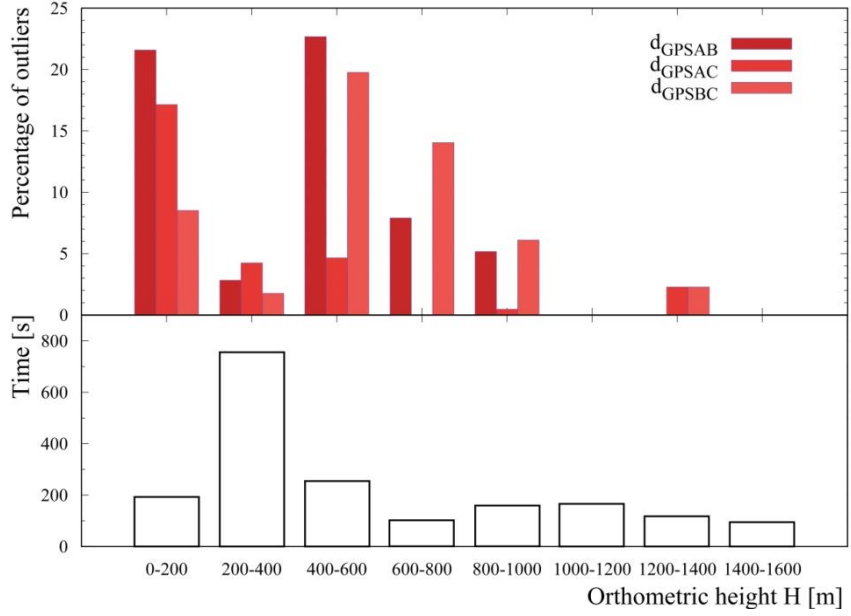

**Figure 8.** In the upper panel are shown the percentages of outliers identified in the d_GPSAB, d_GPSAC and d_GPSBC datasets as function of the orthometric height H. In the bottom panel the acquisition statistics is as function of the orthometric height.

*2.4. The two Pressure and Temperature Sensors*

The Toradex Oak USB Atmospheric Pressure (PT) sensor, hosted inside the Radgyro fuselage (Figure 3), acquires pressure and temperature data with a sampling frequency of 2 Hz and a 10 Pa resolution (corresponding to approximately 0.8 m in height) and a ±2 °C accuracy, respectively.

The pressure and temperature datasets provided by both the PT and the PTIMU devices have been processed by applying the hypsometric formula, which allows estimating the orthometric heights $H_{PT}$ and $H_{PTIMU}$ on the basis of the decreasing exponential trend of the atmospheric pressure with respect to the altitude and accounting for the tiny variations of the temperature in the lower atmosphere:

$$H_{PT} = \frac{T_0}{L}\left[\left(\frac{P(H)}{P_0}\right)^{-\frac{LR}{g}} - 1\right] \qquad (1)$$

where $g = 9.8230$ m/s² is calculated over the flight area at ground level according to the GOCE-based geopotential model described in [26], $R = 287.053$ J/(kg·K) (gas constant for air), $T_0$ is the temperature at sea level (K), $P_0$ is the pressure at the sea level (Pa) and $L = \Delta T/\Delta H = -6.5 \times 10^{-3}$ K/m (temperature lapse rate), constant below 11 km orthometric height [27].

Thanks to the fact that PT and PTIMU are located respectively inside the Radgyro fuselage and inside one lateral compartment (Figure 3), it has been possible to investigate how the Radgyro dynamics affects the pressure readout of both devices, which can be influenced by variations in the air fluxes, by the aircraft velocity as well as by depressions caused by the rotor motion, especially during the take off stage (Figure 9).

Barometric altimeters are not able to provide absolute height without prior knowledge of the local sea level pressure $P_0$. A calibration of the pressure at sea level $P_0$ is necessary in order to take into account the variation of air fluxes related to the Radgyro dynamics as well as possible variations of the atmospheric conditions during the flight [28]. The calibration of PT and PTIMU has been performed applying the inverse hypsometric formula (Equation (1)), where $H_{PT}$ is obtained by averaging the heights measured by GNSS receivers and ALT (at altitude less than 340 m) during 120 s of flight. This interval is chosen on the base of general agreement among sensor data, minimizing the standard deviations during the flight. Since F11 and F14 are characterized by longer acquisition times, this process has been applied during the flight in two different separated intervals.

After these calibrations an internal consistency check shows that all systematic discrepancies of altitude measured by PT and PTIMU have been removed. The successful correction is confirmed by the excellent agreement between $H_{GPSABC}$ and $H_{PT}$ data (Figure 10).

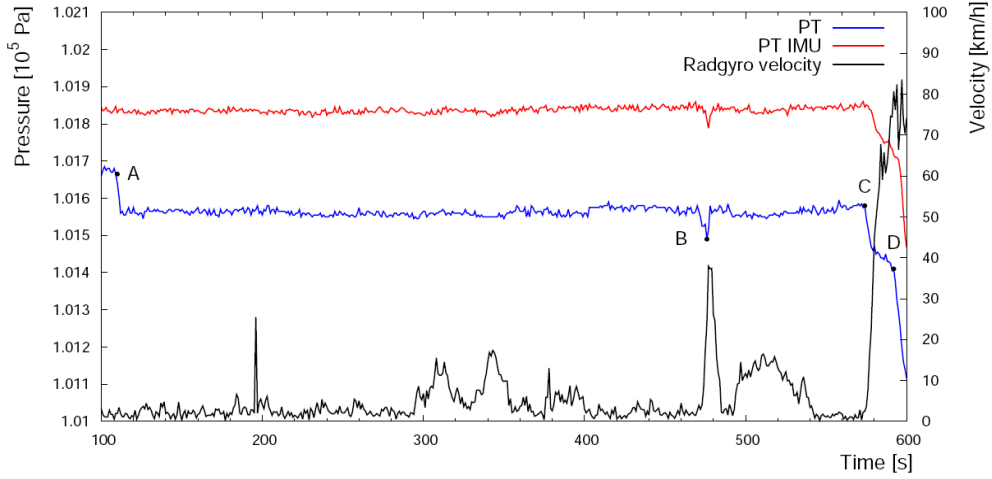

**Figure 9.** Temporal profile of the pressure measured by PT (in blue) and PTIMU (in red) not calibrated, and of the Radgyro horizontal velocity (in black) before the take-off. When the back screw is turned on, the PT sensor, which is significantly exposed to the air flux, measures a depression (point A). The pressure variation registered by both sensors in B is due to the rapid increase of velocity during the taxiing. The accelerating run along a runway starts in C and in D the aircraft takes off.

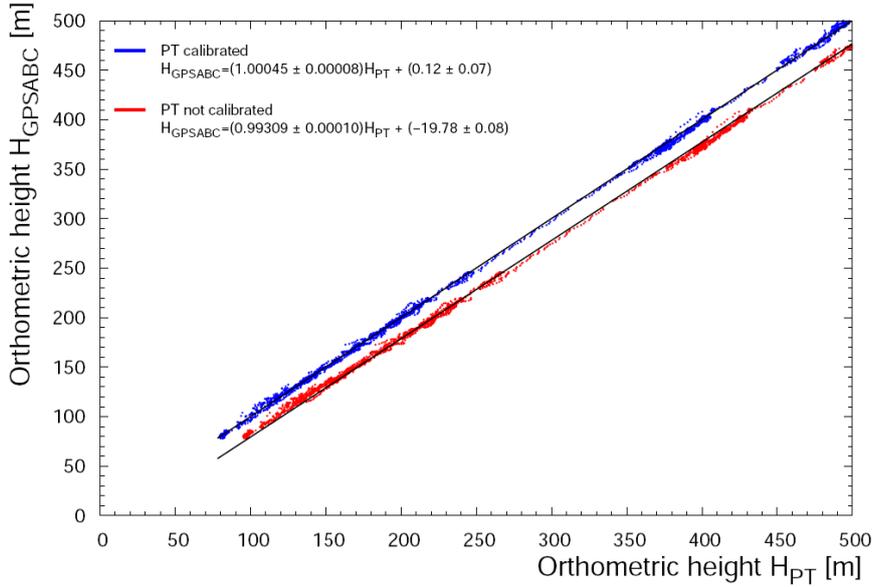

**Figure 10.** Linear regression between $H_{GPSABC}$ and $H_{PT}$ data for F11. In blue and red are reported the calibrated and not-calibrated barometric data respectively. The black straight lines are the linear fits to data: in both cases $r^2 = 0.999$.

## 3. Results and Discussion

This section discusses the comparison and the accuracy of the orthometric heights derived by GNSS, radar altimeter and barometers. The metric adopted is based on the root mean square RMS($\delta H^J$) of the discrepancy between $H^J$ measured by the $J$-th sensor and the averaged height obtained from all the sensors:

$$RMS\left(\delta H^J\right) = \sqrt{\frac{\sum_{i=1}^{N}\left(\delta H_i^J\right)^2}{N}} \qquad (2)$$

where N is the total numbers of data. For each $i$-th measurement obtained by $J$-th sensor, the residual around the mean $\delta H_i^J$ is given by:

$$\delta H_i^J = H_i^J - \bar{H}_i \qquad (3)$$

where $\bar{H}_i$ is the average height measured by M sensors (For H < 340 m, M = 7, while H > 340 m, M = 6, due to the fact the number of outliers of radar altimeter exceeds 30%.):

$$\bar{H}_i = \frac{\sum_{J=1}^{M} H_i^J}{M} \quad (4)$$

Since we have shown that the nominal accuracy of each sensor is often underestimated and we can't hypothesize a priori a quality ranking, we don't introduce any weights in Equation (4). The correlations among the uncertainties of three classes of sensors (GNNS, radar altimeter and barometric sensors) are negligible, while we can't exclude any possible correlations between data acquired by GNNS sensors or barometers. The analysis of the latter correlations goes beyond the purpose of this study, which investigates the main sources of systematic errors. Moreover, Comparing RMS($\delta H^J$) with the mean of the residuals of the *J*-th sensor:

$$\overline{\delta H}^J = \frac{\sum_{i=1}^{N} \delta H_i^J}{N} \quad (5)$$

it is possible to highlight potential systematic biases related to the *J*-th sensor dataset, which can be distinguished from the variance of the residuals $\sigma^2(H^J)$ on the base of the following relationship:

$$RMS(\delta H^J) = \sqrt{\sigma^2(H^J) + \left(\overline{\delta H}^J\right)^2} \quad (6)$$

where $\sigma^2(H^J)$ is defined as:

$$\sigma^2(H^J) = \frac{\sum_{i=1}^{N} \left(\delta H_i^J - \overline{\delta H}^J\right)^2}{N-1} \quad (7)$$

This metric allows emphasizing the systematic height shift related to a specific sensor dataset with respect to the dispersion of values around the average height (see Table 3 and 4). In other words from Equation (6) we learn that when the RMS($\overline{\delta H}^J$) ≈ $|\overline{\delta H}^J|$ then $\sigma(H^J) \ll |\overline{\delta H}^J|$ i.e., the measurement is affected by a systematic bias which dominates the dispersion of data $\sigma(H^J)$. With the perspective of evaluating an overall uncertainty on the height measurement for airborne gamma ray applications, we also calculate the distribution of standard deviations $\sigma_i(H)$ for each *i*-th entry of the dataset:

$$\sigma_i(H) = \sqrt{\frac{\sum_{J=1}^{M} \left(H_i^J - \bar{H}_i\right)^2}{M-1}} \quad (8)$$

The analysis has been performed on four different datasets (Table 2), which have been distinguished according to a spatial selection cut and a GPS processing method cut, corresponding respectively to the 340 m altimeter outlier cutoff and to the 0.2 Hz frequency double-differences dataset of GPSA, GPSB and GPSC post-processing, as described in Sections 2.2 and 2.3.

**Table 3.** Average residuals $\overline{\delta H}^J$ and RMS($\delta H^J$) for data acquired at 1 Hz in the range (35–340) m (DATASET 1α) and in the range (340–2194) m (DATASET 1β).

| | DATASET 1α | | | | | | | | | | | | | |
|---|---|---|---|---|---|---|---|---|---|---|---|---|---|---|
| | GPSA [m] | | GPSB [m] | | GPSC [m] | | GPSIMU [m] | | ALT [m] | | PTIMU [m] | | PT [m] | |
| | $\overline{\delta H}$ | RMS | $\overline{\delta H}$ | RMS | $\overline{\delta H}$ | RMS | $\overline{\delta H}$ | RMS | $\overline{\delta H}$ | RMS | $\overline{\delta H}$ | RMS | $\overline{\delta H}$ | RMS |
| **F11** | −0.1 | 1.8 | 0.7 | 2.7 | 0.4 | 1.9 | 0.0 | 1.7 | 0.0 | 1.5 | −0.8 | 1.7 | −0.2 | 1.4 |
| **F12** | −0.2 | 1.8 | −0.1 | 2.1 | 0.2 | 2.3 | 0.8 | 1.4 | −0.7 | 2.9 | 0.0 | 1.9 | 0.1 | 2.0 |
| **F15** | 1.9 | 2.3 | 0.5 | 2.1 | 1.7 | 2.5 | 5.8 | 5.9 | −3.2 | 3.3 | −4.1 | 4.3 | −2.7 | 3.0 |

| | DATASET 1β | | | | | | | | | | | | |
|---|---|---|---|---|---|---|---|---|---|---|---|---|---|
| | GPSA [m] | | GPSB [m] | | GPSC [m] | | GPSIMU[m] | | ALT [m] | | PTIMU [m] | | PT [m] | |
| | $\overline{\delta H}$ | RMS | $\overline{\delta H}$ | RMS | $\overline{\delta H}$ | RMS | $\overline{\delta H}$ | RMS | $\overline{\delta H}$ | RMS | $\overline{\delta H}$ | RMS | $\overline{\delta H}$ | RMS |
| **F11** | 0.4 | 2.5 | 0.6 | 2.1 | 1.3 | 2.1 | -1.4 | 2.3 | / | / | −0.8 | 2.0 | −0.1 | 1.6 |
| **F14** | 0.7 | 1.7 | 1.0 | 2.0 | 1.5 | 2.2 | -3.1 | 3.4 | / | / | −0.2 | 1.5 | −0.1 | 1.7 |

**Table 4.** Average residuals $\overline{\delta H}^J$ and $RMS(\delta H^J)$ for DATASET 2.

| | GPSA [m] | | GPSB [m] | | GPSC [m] | | GPSIMU[m] | | ALT[m] | | PTIMU [m] | | PT [m] | |
|---|---|---|---|---|---|---|---|---|---|---|---|---|---|---|
| **DATASET 2α** | | | | | | | | | | | | | | |
| | $\overline{\delta H}$ | RMS | $\overline{\delta H}$ | RMS | $\overline{\delta H}$ | RMS | $\overline{\delta H}$ | RMS | $\overline{\delta H}$ | RMS | $\overline{\delta H}$ | RMS | $\overline{\delta H}$ | RMS |
| **F11** | −0.5 | 1.9 | 0.6 | 1.8 | −0.2 | 1.9 | 0.2 | 1.9 | 0.2 | 1.3 | −0.5 | 1.8 | 0.1 | 1.4 |
| **F12** | −0.2 | 1.7 | 0.0 | 1.4 | 0.2 | 1.6 | 1.1 | 1.5 | −0.5 | 2.5 | 0.2 | 1.6 | −0.8 | 1.8 |
| **F15** | 2.1 | 2.4 | 0.0 | 1.7 | 0.4 | 1.1 | 6.8 | 6.9 | −2.4 | 2.5 | −3.8 | 4.1 | −3.1 | 3.6 |
| **DATASET 2β** | | | | | | | | | | | | | | |
| | GPSA [m] | | GPSB [m] | | GPSC [m] | | GPSIMU[m] | | ALT[m] | | PTIMU [m] | | PT [m] | |
| | $\overline{\delta H}$ | RMS | $\overline{\delta H}$ | RMS | $\overline{\delta H}$ | RMS | $\overline{\delta H}$ | RMS | $\overline{\delta H}$ | RMS | $\overline{\delta H}$ | RMS | $\overline{\delta H}$ | RMS |
| **F11** | 0.1 | 2.3 | 0.6 | 1.3 | 0.9 | 1.7 | −0.1 | 1.3 | / | / | −1.4 | 2.4 | −0.1 | 1.6 |
| **F14** | 0.6 | 1.3 | 0.4 | 1.3 | 0.6 | 1.3 | −1.6 | 2.0 | / | / | −0.1 | 1.3 | 0.1 | 1.7 |

*3.1. Analysis of DATASET 1*

The main results of the analysis of the 1 Hz stand-alone DATASET 1α and 1β (Table 2) are summarized for each sensor in Table 3 in terms of average of the residuals $\overline{\delta H}^J$ and of the root mean square of the residuals $RMS(\delta H^J)$. The poor accuracy of data from GNSS at low altitude mentioned in Section 2.3 is confirmed in this analysis. In particular, for H < 66 m (F15) we note not only a dispersion of values, but also a clear systematic shift of altitude measured by this class of sensors with respect to those obtained by radar and barometric altimeters. This evidence shows that the multipath effect at low altitude produces severe interferences for low-cost GNSS receivers.

In the (79–340) m range the agreement among values of altitude measured by all seven sensors is good (Figure 11). The median of the distribution of standard deviations is 1.7 m and the values of $\overline{\delta H}^J$ reported in Table 3 do not highlight any significant systematic effect (i.e., $\overline{\delta H}^J$ < 1 m). Finally in the (340–2194) m range the median of the distribution of standard deviations is 2.1 m. The comparison of different values of $\overline{\delta H}^J$ calculated for GNSS receivers seems to show two clusters of data characterized by positive and negative values of $\overline{\delta H}^J$ for GPSABC and GPSIMU respectively. This feature is evident in Figure 12 where the values closest to zero for PT and PTIMU highlight how these barometric sensors give the best performance when they are calibrated with redundant measurements from GNSS receivers.

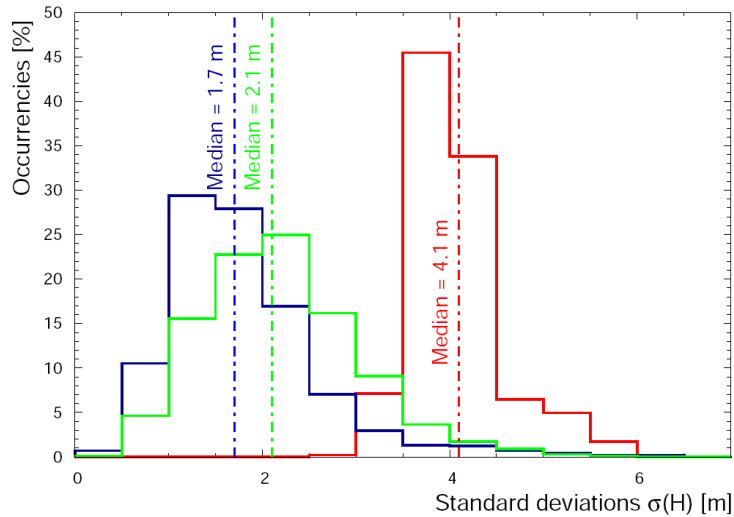

**Figure 11.** Distribution of σ(H) (standard deviations of heights) in the range (35–66) m (red solid line), (79–340) m (blue solid line) and (340–2194) m (green solid line) measured at 1 Hz.

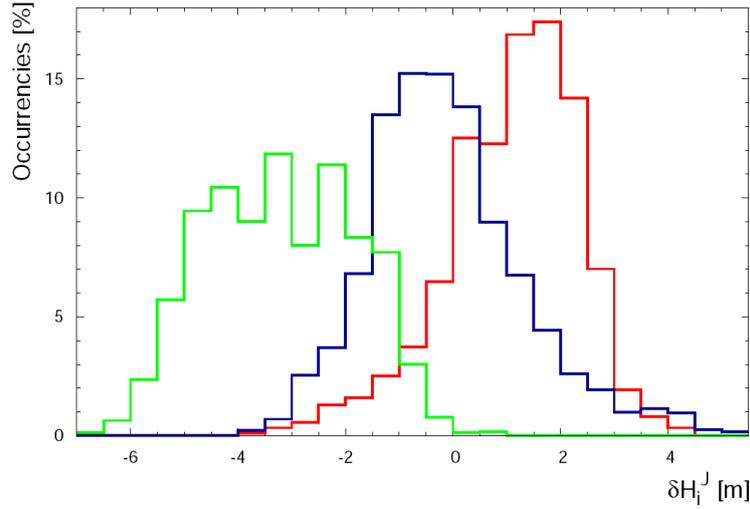

**Figure 12.** Distribution of the residuals $\delta H_i^J$ in the (464–2194) m range of altitude: GPSIMU dataset in solid green line, PTIMU and PT dataset is reported in solid blue line, and GPSABC dataset in solid red line.

*3.2. Analysis of DATASET 2*

In this section the analysis of the 0.2 Hz double-difference DATASET 2α and DATASET 2β is presented, in which the GPSA, GPSB and GPSC acquisitions have been post-processed using the double differences method with respect to the master station observation data. PT and PTIMU are calibrated with the criteria described in Section 2.4 using the GNSS double-difference data.

The double differences post-processing increases the quality of GPSA, GPSB and GPSC data which reflects in a shrinking of the $\sigma_i(H)$ distribution towards low standard deviation values at altitude (79–2194) m (Figure 13 Panel b), making the median value of the distribution decrease from 1.5 m (stand-alone) to 0.8 m (double-differences). We note that the major benefit of the data treatment affects the GPSABC accuracy at high altitude [340–2194] m, where the values of $\overline{\delta H}^J$ are comparable to those obtained for (79–340) m (i.e., $\overline{\delta H}^J < 1$ m) (Table 4).

On the basis of $\overline{\delta H}^J$ and RMS($\delta H^J$) calculated in Table 4 we can assert that double differences post-processing does not produce any evident improvement of the altitude accuracy at H < 66 m (Figure 13 Panel a). The severe multipath noise which affects the calculation of relative distances between the three GPS antennas and the altitude measured at 1 Hz, is not healed by double-differences post processing at heights lower than 66 m. On the other hand, the non-GNSS sensors give excellent results in terms of linear correlation and negligible systematic effects. Observing linear regression data in Table A3 for PT, PTIMU, and ALT the slope and the intercept are compatible with 1 and 0 at 2σ-level respectively.

The performance of all sensors in the altitude range (79–2194) m is essentially similar: the RMS($\delta H^J$) varies from 1.3 m to 2.5 m and the maximum $\overline{\delta H}^J = 1.6$ m (Table 4). The distribution of standard deviation of heights reported in Figure 14 does not show any peculiar feature: for altitude ranges of (79–340) m and (340–2194) m the medians are 1.6 m and 1.5 m, respectively. In particular the Pearson correlation coefficients calculated for all the couples of sensors highlight perfect linear correlation in the (340–2194) m range of altitude (Tables A4 and A5).

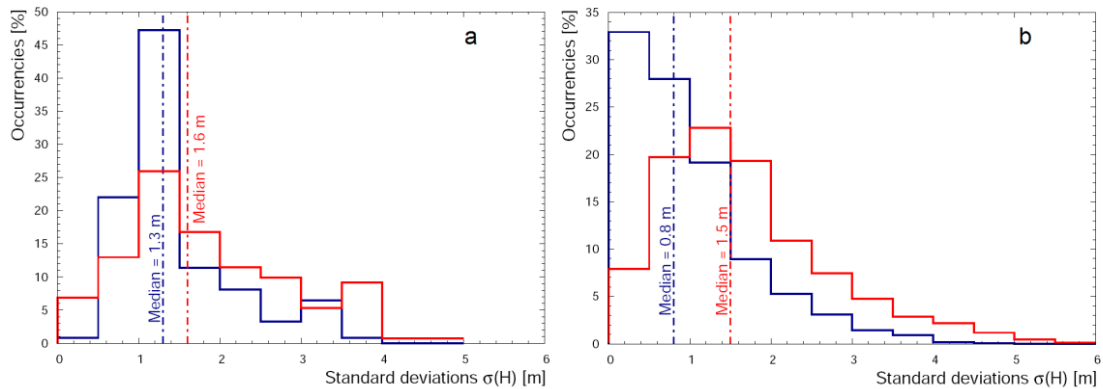

**Figure 13.** Distribution of σ(H) (standard deviations of heights) calculated for GPSABC built-in solution (red solid line) and with double-difference post-processing (blue solid line), in the altitude ranges (35–66) m (**panel a**) and (79–2194) m (**panel b**).

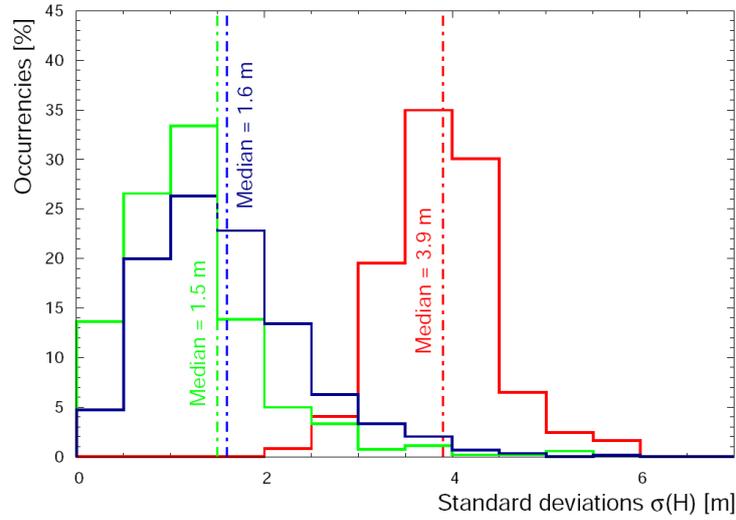

**Figure 14.** Distribution of σ(H) (standard deviations of heights) in the altitude ranges (35–66) m (red solid line), (79–340) m (blue solid line) and (340–2194) m (green solid line) measured at 0.2 Hz.

*3.3. Effect of the Accuracy of the Flight Altitude on AGRS Measurements*

In airborne gamma-ray spectroscopy measurements, knowing the survey altitude above the ground is of crucial importance in order to model the exponential attenuation photons having different characteristic energies suffer when traversing the air material. This aspect has implications on the determination of the height correction factors which are separately calculated for each radionuclide in order to reconstruct the counting statistics at ground level in case of flat morphology.

In fact, neglecting possible systematic uncertainties originating from the calibration of the instrumental setup, the main source of uncertainty in AGRS measurements is related to the counting statistics, which result from the statistical nature of both radioactive decay and photon attenuation in traversing materials. Considering a NaI(Tl) detector having a volume of 16 L and flying at a height of 100 m over a terrain characterized by K, U and Th abundances respectively equal to 0.02 g/g, 2.5 μg/g and 9.0 μg/g, the statistics in counts per second (cps), recorded in the three energy windows typically used to quantify the content of the three radionuclides (1370–1570 keV for K, 1660–1860 keV for U and 2410–2810 keV for Th), are respectively (103.2 ± 10.2) cps, (12.4 ± 3.5) cps and (26.5 ± 5.1) cps, where the uncertainty is estimated according to the Poisson distribution describing the nature of the radioactive decay.

Assuming that the detector receives the signal produced by a homogeneous infinite half-space soil volume source, the measured count rate can be written as (see Appendix C):

$$N(E,z) = N(E,0) \int_0^1 d\cos\theta \, e^{-\frac{\mu(E)z}{\cos\theta}} \tag{9}$$

where $N(E,z)$ and $N(E,0)$ are respectively the counting statistics in counts per second [cps] at the survey altitude $z$ and the one that would have been measured at ground level for photons having emission energy $E$, and $\mu(E)$ [m$^{-1}$] is the air linear attenuation coefficient, corresponding to the inverse of the mean free path traveled by a photon having energy $E$ and traversing the air material. As the $^{40}$K, $^{238}$U, $^{232}$Th ground abundances are determined by dividing the estimated ground level count rates $N(E,0)$ in each specific energy window by the corresponding ground sensitivity constants (i.e., the count rate per unit radioisotope concentration), the relative uncertainty on the ground abundance is the same affecting the counting statistics $N(E,0)$, assuming that the uncertainty on the sensitivity constants is negligible.

In order to establish the effect of the height uncertainty on the radioactive content assessment, it is necessary to go through an uncertainty propagation procedure, according to which the relative uncertainty on the ground counting statistics $\frac{\sigma_{N(E,0)}}{N(E,0)}$ can be written as follows (see Appendix C):

$$\left(\frac{\sigma_{N(E,0)}}{N(E,0)}\right)^2 = \left(\frac{\sigma_{N(E,z)}}{N(E,z)}\right)^2 + (\beta(\mu(E),z)\mu(E)\sigma_z)^2$$

where:

$$\beta(\mu(E),z) = \frac{\int_0^1 dcos\theta \frac{1}{cos\theta} e^{\frac{-\mu(E)z}{cos\vartheta}}}{\int_0^1 dcos\theta \, e^{\frac{-\mu(E)z}{cos\vartheta}}} \quad (10)$$

and $\frac{\sigma_{N(E,z)}}{N(E,z)}$ and $\sigma_z$ are respectively the relative uncertainty on the measured counting statistics and the absolute uncertainty on the height above ground level. The adimensional $\beta(\mu(E),z)$ function has a lower limit value equal to 1 and has a weak dependence on the photon energy and a stronger dependence on the survey altitude, as can be seen from Figure A1.

Assuming that the uncertainty on the height above ground level is exclusively given by the uncertainty on the orthometric height (i.e., neglecting any contribution that can potentially arise from the Digital Elevation Model (DEM) of the terrain) we estimate by choosing as $\sigma_z$ the 2.1 m median value of the standard deviation distribution of DATASET 1β that the relative uncertainty on the $^{40}$K, $^{238}$U, $^{232}$Th ground abundances at 100 m are 2.2%, 2.0% and 1.7% respectively, which result only from the uncertainty on the height above the ground (i.e., neglecting the uncertainty on the counting statistics).

## 4. Conclusions

In this study we investigate how the combination of redundant data from seven altimetric sensors can improve the quality of flight height accuracy with the perspective to estimate the uncertainties affecting airborne radiometric measurements. In flying over the sea in the range (35–66) m, the altitude measured by GNSS antennas suffers the severe noise due to the multipath effect which is clearly reduced on the land. Although the IMU sensor provides the flight altitude by combining the data from the GNSS, the barometer, and the accelerometers, the values of $\overline{\delta H}^{IMU}$ and RMS($\delta H^{IMU}$) are greater than 5 m, highlighting how the external correction is not effective in mitigating the noise due to the multipath on the GPS signal.

On the base of our study, we can conclude that the most accurate measurement of flight altitude over the sea in the (35–66) m range has been performed by two barometric altimeters together with the radar altimeter. In this altitude regime the linear regressions (Table A3) show slopes and intercepts compatible with 1 and 0 at 2σ-level, and the median of the distribution of standard deviations of heights is 1.1 m. Adopting this altitude error at 50 m, the relative uncertainties on the $^{40}$K, $^{238}$U, $^{232}$Th ground abundances are equal to 1.3%, 1.2% and 1.1% respectively.

According to our investigation the reliability of radar altimeter is in agreement with its declared accuracy (3% of the altitude value) up to 340 m: beyond this height the number of outliers increases drastically, preventing the inclusion of these data for the comparative analysis. In the (79–340) m range the median of the distribution of standard deviations of altitude acquired by all seven sensors is 1.6 m, with double differences post-processing of the signal recorded by three single frequency GNSS antennas.

Adopting conservatively this uncertainty for a 100 m flight altitude, we can estimate a relative uncertainty of 1.7%, 1.5% and 1.3% associated to $^{40}$K, $^{238}$U, $^{232}$Th ground abundances respectively. At altitude higher than 79 m, the GNSS double-difference post-processing enhanced significantly the data quality obtained by the 3 low-cost and light antennas. This is proved by a reduction of the median value of the standard deviations (from 1.5 m with stand-alone analysis to 0.8 m with double-difference processing) and by an increasing precision in the reconstruction of median distance of the three antennas with increasing altitude. Since the computation of double differences does not solve the multipath problem, the use of better performing antennas with size and cost compatible with AGRS survey is strongly recommended.

According to our study, the best integration of data from GNNS antennas, radar and barometric altimeters allows reaching an accuracy better than 2% at flight altitude higher than ~80 m, which affect the estimation of ground total activity measured at 100 m with an uncertainty resulting from the sole uncertainty on the flight height of the order of 2%.

**Acknowledgments:** This work was partially founded by the National Institute of Nuclear Physics (INFN) through the ITALian RADioactivity project (ITALRAD) and by the Theoretical Astroparticle Physics (TAsP) research network. The co-authors would like to acknowledge the support of the Geological and Seismic Survey of the Umbria Region (UMBRIARAD), of the University of Ferrara (Fondo di Ateneo per la Ricerca scientifica FAR 2016), of the Project Agroalimentare Idrointelligente CUP D92I16000030009 and of the MIUR (Ministero dell'Istruzione, dell'Università e della Ricerca) under MIUR-PRIN-2012 project. The authors would like to thank the staff of GeoExplorer Impresa Sociale s.r.l. for their support and Mauro Antogiovanni, Claudio Pagotto, Ivan Callegari and Andrea Motti for their collaboration which made possible the realization of this study. The authors would also like to show their gratitude to Emanuele Tufarolo for useful discussions.

**Appendix A**

For the flights with data under 340 m are shown the values of the linear regressions between each pair of sensors in the DATASET 2α: m (slope) and q (y-intercept) with their uncertainties and the $r^2$ (correlation coefficient). The sensors on the first row of each table give us the x values and the sensors on the first column give the y values.

**Table A1.** Linear regression data of F11 (DATASET 2α).

| | | F11 | | | | | |
|---|---|---|---|---|---|---|---|
| | | GPSB | GPSA | GPSIMU | ALT | PT | PTIMU |
| **GPSC** | m | 0.994 ± 0.002 | 0.992 ± 0.003 | 0.981 ± 0.003 | 0.987 ± 0.002 | 0.996 ± 0.003 | 1.001 ± 0.003 |
| | q | 0.21 ± 0.38 | −1.56 ± 0.49 | 2.62 ± 0.58 | 1.72 ± 0.39 | 0.48 ± 0.51 | 0.05 ± 0.55 |
| | $r^2$ | 0.998 | 0.997 | 0.996 | 0.998 | 0.997 | 0.996 |
| **GPSB** | m | | 0.998 ± 0.003 | 0.987 ± 0.003 | 0.993 ± 0.002 | 1.001 ± 0.003 | 1.007 ± 0.003 |
| | q | | 1.46 ± 0.44 | 2.49 ± 0.51 | 1.63 ± 0.33 | −0.43 ± 0.52 | 0.00 ± 0.56 |
| | $r^2$ | | 0.998 | 0.997 | 0.999 | 0.997 | 0.996 |
| **GPSA** | m | | | 0.988 ± 0.003 | 0.993 ± 0.002 | 1.001 ± 0.003 | 1.008 ± 0.003 |
| | q | | | 1.28 ± 0.545 | 0.40 ± 0.36 | −0.85 ± 0.49 | −1.31 ± 0.51 |
| | $r^2$ | | | 0.996 | 0.998 | 0.997 | 0.997 |
| **GPSIMU** | m | | | | 1.003 ± 0.003 | 1.012 ± 0.002 | 1.019 ± 0.002 |
| | q | | | | −0.48 ± 0.46 | −1.95 ± 0.37 | −2.42 ± 0.38 |
| | $r^2$ | | | | 0.997 | 0.998 | 0.998 |
| **ALT** | m | | | | | 1.008 ± 0.002 | 1.014 ± 0.003 |
| | q | | | | | −1.20 ± 0.40 | −1.63 ± 0.46 |
| | $r^2$ | | | | | 0.998 | 0.997 |
| **PT** | m | | | | | | 1.007 ± 0.001 |
| | q | | | | | | 0.44 ± 0.19 |
| | $r^2$ | | | | | | 1.000 |

**Table A2.** Linear regression data of F12 (DATASET 2α).

| | | F12 | | | | | |
|---|---|---|---|---|---|---|---|
| | | GPSB | GPSA | GPSIMU | ALT | PT | PTIMU |
| **GPSC** | m | 1.016 ± 0.005 | 0.996 ± 0.006 | 1.015 ± 0.006 | 1.056 ± 0.007 | 1.015 ± 0.007 | 0.998 ± 0.007 |
| | q | −2.67 ± 0.81 | 1.15 ± 1.01 | −3.65 ± 1.01 | −9.25 ± 1.27 | −1.70 ± 1.25 | 0.28 ± 1.21 |
| | $r^2$ | 0.997 | 0.995 | 0.995 | 0.993 | 0.993 | 0.993 |
| **GPSB** | m | | 0.980 ± 0.004 | 0.998 ± 0.005 | 1.037 ± 0.008 | 0.997 ± 0.007 | 0.981 ± 0.007 |
| | q | | 3.80 ± 0.70 | −0.77 ± 0.90 | −6.08 ± 1.36 | 1.25 ± 1.25 | 3.18 ± 1.19 |
| | $r^2$ | | 0.998 | 0.996 | 0.992 | 0.993 | 0.993 |
| **GPSA** | m | | | 1.017 ± 0.005 | 1.057 ± 0.007 | 1.016 ± 0.008 | 1.000 ± 0.007 |
| | q | | | −4.43 ± 0.95 | −9.93 ± 1.33 | −2.27 ± 1.38 | −0.35 ± 1.28 |
| | $r^2$ | | | 0.996 | 0.993 | 0.991 | 0.992 |
| **GPSIMU** | m | | | | 1.036 ± 0.008 | 1.000 ± 0.004 | 0.983 ± 0.003 |
| | q | | | | −4.90 ± 1.42 | 1.93 ± 0.75 | 3.85 ± 0.63 |
| | $r^2$ | | | | 0.991 | 0.997 | 0.998 |
| **ALT** | m | | | | | 0.957 ± 0.008 | 0.941 ± 0.008 |
| | q | | | | | 7.96 ± 1.38 | 9.92 ± 1.41 |
| | $r^2$ | | | | | 0.99 | 0.99 |
| **PT** | m | | | | | | 0.983 ± 0.002 |
| | q | | | | | | 2.07 ± 0.44 |
| | $r^2$ | | | | | | 0.999 |

Table A3. Linear regression data of F15 (DATASET 2α).

| | | GPSB | GPSA | GPSIMU | ALT | PT | PTIMU |
|---|---|---|---|---|---|---|---|
| **GPSC** | m | 0.896 ± 0.017 | 0.956 ± 0.008 | 1.015 ± 0.014 | 0.982 ± 0.007 | 0.991 ± 0.027 | 1.022 ± 0.026 |
| | q | 5.68 ± 0.89 | 0.65 ± 0.43 | −7.34 ± 0.82 | 3.65 ± 0.35 | 3.92 ± 1.32 | 3.14 ± 1.25 |
| | $r^2$ | 0.958 | 0.992 | 0.978 | 0.994 | 0.928 | 0.962 |
| **GPSB** | m | | 1.027 ± 0.019 | 1.108 ± 0.016 | 1.057 ± 0.019 | 1.079 ± 0.031 | 1.113 ± 0.030 |
| | q | | −3.47 ± 1.05 | −13.05 ± 0.95 | −0.34 ± 0.92 | −0.63 ± 1.50 | −1.51 ± 1.41 |
| | $r^2$ | | 0.958 | 0.975 | 0.964 | 0.968 | 0.921 |
| **GPSA** | m | | | 1.054 ± 0.017 | 1.022 ± 0.008 | 1.032 ± 0.028 | 1.061 ± 0.028 |
| | q | | | −7.89 ± 0.97 | 3.37 ± 0.40 | 3.36 ± 1.38 | 2.97 ± 1.35 |
| | $r^2$ | | | 0.971 | 0.993 | 0.916 | 0.922 |
| **GPSIMU** | m | | | | 0.952 ± 0.11 | 0.970 ± 0.025 | 1.001 ± 0.023 |
| | q | | | | 11.57 ± 0.54 | 11.38 ± 1.21 | 10.57 ± 1.11 |
| | $r^2$ | | | | 0.984 | 0.927 | 0.939 |
| **ALT** | m | | | | | 1.009 ± 0.027 | 1.040 ± 0.026 |
| | q | | | | | 0.33 ± 1.31 | −0.45 ± 1.24 |
| | $r^2$ | | | | | 0.921 | 0.931 |
| **PT** | m | | | | | | 1.020 ± 0.009 |
| | q | | | | | | −0.28 ± 0.45 |
| | $r^2$ | | | | | | 0.990 |

**Appendix B**

For the flights with data over 340 m are shown the values of the linear regressions between each pair of sensors in the DATASET 2β: m (slope) and q (y-intercept) with their uncertainties and the $r^2$ (correlation coefficient). The sensors on the first row of each table give us the x values and the sensors on the first column give the y values.

Table A4. Linear regression data of F11 (DATASET2β).

| | | GPSB | GPSA | GPSIMU | PT | PTIMU |
|---|---|---|---|---|---|---|
| **GPSC** | m | 0.9996 ± 0.0001 | 1.0007 ± 0.0001 | 1.0009 ± 0.0002 | 0.9994 ± 0.0002 | 0.9987 ± 0.0002 |
| | q | 0.75 ± 0.11 | 0.11 ± 0.14 | 0.15 ± 0.21 | 1.57 ± 0.24 | 3.50 ± 0.26 |
| | $r^2$ | 1.000 | 1.000 | 1.000 | 1.000 | 1.000 |
| **GPSB** | m | | 1.0011 ± 0.0002 | 1.0013 ± 0.0001 | 0.9998 ± 0.0002 | 0.9992 ± 0.0002 |
| | q | | −0.64 ± 0.18 | −0.60 ± 0.16 | 0.82 ± 0.21 | 2.75 ± 0.23 |
| | $r^2$ | | 1.000 | 1.000 | 1.000 | 1.000 |
| **GPSA** | m | | | 1.0002 ± 0.0002 | 0.9987 ± 0.0003 | 0.9980 ± 0.0003 |
| | q | | | 0.05 ± 0.28 | 1.48 ± 0.30 | 3.40 ± 0.33 |
| | $r^2$ | | | 1.000 | 1.000 | 1.000 |
| **GPSIMU** | m | | | | 0.9985 ± 0.0001 | 0.9979 ± 0.0001 |
| | q | | | | 1.41 ± 0.12 | 3.34 ± 0.11 |
| | $r^2$ | | | | 1.000 | 1.000 |
| **PT** | m | | | | | 0.9994 ± 0.0001 |
| | q | | | | | 1.93 ± 0.11 |
| | $r^2$ | | | | | 1.000 |

Table A5. Linear regression data of F14 (DATASET2β).

| | | F14 | | | | |
|---|---|---|---|---|---|---|
| | | GPSB | GPSA | GPSIMU | PT | PTIMU |
| **GPSC** | m | 0.99997 ± 0.00005 | 0.9996 ± 0.0001 | 0.9977 ± 0.0001 | 0.9975 ± 0.0002 | 0.9982 ± 0.0002 |
| | q | 0.18 ± 0.08 | 0.62 ± 0.10 | 5.64 ± 0.20 | 4.25 ± 0.26 | 3.37 ± 0.24 |
| | $r^2$ | 1.000 | 1.000 | 1.000 | 1.000 | 1.000 |
| **GPSB** | m | | 0.99960 ± 0.00004 | 0.9977 ± 0.0001 | 0.9976 ± 0.0002 | 0.9982 ± 0.0002 |
| | q | | 0.44 ± 0.06 | 5.47 ± 0.22 | 4.07 ± 0.28 | 3.19 ± 0.25 |
| | $r^2$ | | 1.000 | 1.000 | 1.000 | 1.000 |
| **GPSA** | m | | | 0.9981 ± 0.0001 | 0.9980 ± 0.0002 | 0.9986 ± 0.0002 |
| | q | | | 5.03 ± 0.21 | 3.63 ± 0.29 | 2.75 ± 0.25 |
| | $r^2$ | | | 1.000 | 1.000 | 1.000 |
| **GPSIMU** | m | | | | 0.9998 ± 0.0001 | 1.0005 ± 0.0001 |
| | q | | | | −1.40 ± 0.21 | −2.28 ± 0.18 |
| | $r^2$ | | | | 1.000 | 1.000 |
| **PT** | m | | | | | 1.0007 ± 0.0001 |
| | q | | | | | −0.87 ± 0.17 |
| | $r^2$ | | | | | 1.000 |

**Appendix C**

The count rate measured in a given energy window by a detector that is flying at altitude z above ground level and which originates from a homogenous infinite half-space soil volume source is given by the following equation:

$$N(E,z) = \frac{A_\gamma S}{2\mu_s(E)} \int_0^1 d\cos\theta \, e^{\frac{-\mu(E)z}{\cos\theta}} \tag{A1}$$

where $A_\gamma$ is the density of photons isotropically emitted by the homogeneous volume source [γ/m³], $S$ is the detector cross-sectional area [m²], $\mu_s(E)$ [m⁻¹] and $\mu(E)$ [m⁻¹] are the soil and air linear attenuation coefficients, corresponding to the inverse of the mean free path traveled by a photon having energy $E$ and traversing the soil and air materials, respectively. Starting from Equation (A1) it is possible to write the following relation between the count rate measured at altitude $z$ $N(E,z)$ and the count rate that one would have measured by placing the detector on the ground $N(E,0)$:

$$N(E,z) = N(E,0)k(\mu(E),z) \tag{A2}$$

where the function $k(\mu(E),z)$ is given by:

$$k(\mu(E),z) = \int_0^1 d\cos\theta \, e^{-\frac{\mu(E)z}{\cos\theta}} \tag{A3}$$

Gamma-ray spectrometers typically undergo a ground efficiency calibration procedure for the determination of the sensitivity constants, which allow for the conversion of measured count rates into ground radionuclide abundances. The $^{40}$K, $^{238}$U, $^{232}$Th ground abundances are therefore predicted by dividing the estimated ground level count rate $N(E,0)$ by the specific ground sensitivity constant, which represents the count rate per unit radioisotope concentration. In the hypothesis of negligible uncertainty on the sensitivity constants, the relative uncertainty on the ground abundance is equal to the relative uncertainty affecting the counting statistics $N(E,0)$, inferred from the measured $N(E,z)$ according to Equation (A2).

We therefore apply the standard propagation of uncertainty for uncorrelated variables to the quantity $N(E,0)$ in order to estimate the contribution given by the uncertainty on the survey altitude $\sigma_z$ to the ground concentration estimation, which can be expressed according to the following equation:

$$\begin{aligned}\sigma^2_{N(E,0)} &= \left(\left|\frac{\partial N(E,0)}{\partial N(E,z)}\right|\sigma_{N(E,z)}\right)^2 + \left(\left|\frac{\partial N(E,0)}{\partial k(\mu(E),z)}\right|\sigma_{k(\mu(E),z)}\right)^2 \\ &= \left(\left|\frac{\partial N(E,0)}{\partial N(E,z)}\right|\sigma_{N(E,z)}\right)^2 + \left(\left|\frac{\partial N(E,0)}{\partial k(\mu(E),z)}\right|\left|\frac{\partial k(\mu(E),z)}{\partial z}\right|\sigma_z\right)^2\end{aligned} \tag{A4}$$

Starting from the inverse of Equation (A2) it is possible to determine all the terms of Equation (A4) as follows:

$$\left|\frac{\partial N(E,0)}{\partial N(E,z)}\right| = \frac{1}{k(\mu(E),z)} \tag{A5}$$

$$\sigma_{N(E,z)} = \sqrt{N(E,z)} \tag{A6}$$

$$\left|\frac{\partial N(E,0)}{\partial k(\mu(E),z)}\right| = \frac{N(E,z)}{k(\mu(E),z)^2} \tag{A7}$$

$$\left|\frac{\partial k(\mu(E),z)}{\partial z}\right| = \int_0^1 dcos\theta \, \frac{\mu(E)}{cos\theta} e^{-\frac{\mu(E)z}{cos\theta}} \tag{A8}$$

where the absolute uncertainty on the observed counting statistics given by Equation (A6) comes from the Poissonian nature of the radioactive decay process. By combining all the obtained terms into Equation (A4) it is possible to write the relative uncertainty on the inferred ground counting statistics as the sum of two terms as stated in the following equation:

$$\left(\frac{\sigma_{N(E,0)}}{N(E,0)}\right)^2 = \left(\frac{\sigma_{N(E,z)}}{N(E,z)}\right)^2 + (\beta(\mu(E),z)\mu(E)\sigma_z)^2$$

$$= \left(\frac{\sigma_{N(E,z)}}{N(E,z)}\right)^2 + \left(\mu(E)\sigma_z \frac{\int_0^1 dcos\theta \, \frac{1}{cos\theta} e^{-\frac{\mu(E)z}{cos\theta}}}{\int_0^1 dcos\theta \, e^{-\frac{\mu(E)z}{cos\theta}}}\right)^2 \tag{A9}$$

The first term in the summation corresponds to the relative uncertainty on the measured counting statistics, while the second term contains the product of the absolute height uncertainty $\sigma_z$ times the air linear attenuation coefficient $\mu(E)$ which is in turn weighted by an adimensional factor $\beta(\mu(E),z)$ that depends on the linear attenuation coefficient itself and on the absolute survey height $z$. The $\beta(\mu(E),z)$ function contains an energy dependence as the attenuation suffered by photons is stronger for decreasing energy (i.e., the $\mu(E)$ linear attenuation coefficient decreases for increasing energy) and has also a $z$ dependence according to which, for a fixed energy, the value of $\beta(\mu(E),z)$ decreases for increasing height asymptotically approaching 1.

Figure A1 shows the profile of the $\beta(\mu(E),z)$ function for the characteristic emission energies of $^{40}$K (1460 keV), $^{214}$Bi (1765 keV) and $^{208}$Tl (2614 keV), where $^{214}$Bi and $^{208}$Tl are the major gamma emitting radionuclides respectively belonging to the $^{238}$U and $^{232}$Th decay series. The adopted values for the $\mu(E)$ gamma linear attenuation coefficients for 1460 keV, 1765 keV and 2614 keV photons propagating in air are respectively equal to 0.006430 m$^{-1}$, 0.005829 m$^{-1}$ and 0.004717 m$^{-1}$. These values have been estimated using an air density of 1.225 kg/m$^3$ and gamma mass attenuation coefficients taken from the National Institute of Standard and Technology website [27], where an air composition of 78% N$_2$, 21% O$_2$ and 1% Ar by weight has been given for the description of the composite traversed material.

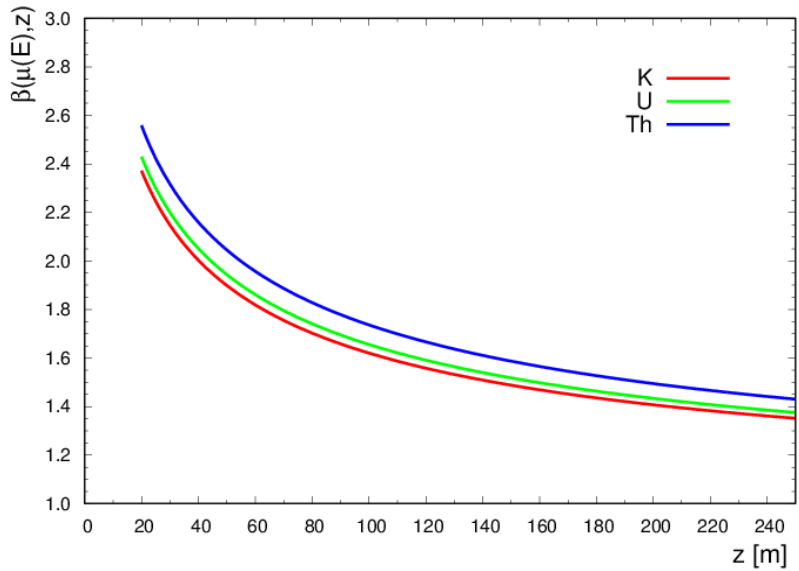

**Figure A1.** Plot of the $\beta(\mu(E),z)$ factor as function of the survey altitude $z$: the red, green and blue line refer respectively to the $^{40}$K, $^{214}$Bi ($^{238}$U) and $^{208}$Tl ($^{232}$Th) gamma emission energies.

According to Equation (A9) it is possible to estimate the relative uncertainty on the K, U and Th abundances due solely to the uncertainty on the survey altitude $\sigma_z$, i.e., for a relative uncertainty on the measured statistics $\left(\frac{\sigma_{N(E,z)}}{N(E,z)}\right) = 0$. As in this study we are not considering the implications related to the use of the Digital Elevation Model (DEM) of the terrain, we adopt as $\sigma_z$ value the median of the distribution of standard deviations of orthometric heights, which is in the worst case scenario (DATASET 1β for (340–2194) m) equal to 2.1 m. Referring to a standard survey height above the ground of 100 m, the estimated relative uncertainties on the K, U and Th concentrations deriving from the uncertainty on the survey height are respectively equal to 2.2%, 2.0% and 1.7%.